\documentstyle[11pt,paspconf,epsf] {article}
\def\wisk#1{\ifmmode{#1}\else{$#1$}\fi}

\def\kms{\ifmmode {\>{\rm km\ s}^{-1}}\else {km s$^{-1}$}\fi}
\def\gtrapprox{\;\lower 0.5ex\hbox{$\buildrel >
    \over \sim\ $}}             
\def\lessapprox{\;\lower 0.5ex\hbox{$\buildrel < \over \sim\ $}}
\def\etal{et al.\ }
\def\eg{e.g.,\ }
\def\ie{i.e.,\ }

\def\psqcm{\ifmmode {\>{\rm cm}^{-2}}\else {cm$^{-2}$}\fi}
\def\pcubcm{\ifmmode {\>{\rm cm}^{-3}}\else {cm$^{-3}$}\fi}
\def\be{\begin{equation}}
\def\ee{\end{equation}}
\def\bea{\begin{eqnarray}}
\def\eea{\end{eqnarray}}
\def\rpcsq{\ifmmode {r_{\rm pc}^2}\else {$r_{\rm pc}^2$}\fi}
\def\rpc{\ifmmode {r_{\rm pc}}\else {$r_{\rm pc}$}\fi}
\def\fabs{\ifmmode {f_{\rm abs}}\else {$f_{\rm abs}$}\fi}
\def\msol{\ifmmode {\>M_\odot}\else {$M_\odot$}\fi}
\def\lsol{\ifmmode {\>L_\odot}\else {$L_\odot$}\fi}
\begin{document}

\title{Physical Conditions in Obscuring Tori and Molecular Accretion
Disks, and Are They Really the Same Thing?}

\author{Philip R. Maloney} 
\affil{Center for Astrophysics and Space
Astronomy, University of Colorado, Boulder, CO 80309-0389}

\begin{abstract}
The nature of the obscuring material in active galactic nuclei is
still uncertain. Although some sources, such as Cygnus A, show evidence
for a geometrically thick ``torus'' as originally suggested, recent
work on the radiation-driven warping instability discovered by Pringle
suggests that obscuration by thin, warped disks may play a crucial
role in AGN.

\end{abstract}


\section{Introduction}

In the currently popular and attractive ``unified model'' for active
galactic nuclei, all AGN are surrounded by a geometrically and
optically thick ``torus'' of obscuring material (see Conway, this
volume, for the proper definition); the orientation of this torus with
respect to our line of sight determines whether the AGN is classified
as a Type I (Seyfert 1 or quasar) or Type II (Sy 2 or radio galaxy)
object. Excellent reviews of unification are given by Antonucci (1993)
and Urry \& Padovani (1995). Following the pioneering theoretical work
of Krolik \& Begelman (1986, 1988), it has been generally assumed that
the tori are composed of dusty molecular clouds; the geometric
thickness of the torus is the consequence of a large cloud velocity
dispersion. How such an assemblage could be maintained (\ie preventing
collisional dissipation of the random cloud motions) is still an
unsolved theoretical problem (Krolik \& Begelman 1988; Pier \& Krolik
1992). However, there is indisputable evidence for the presence of
obscuring material in many Type II AGN, and it is clear that this
plays a major role in the classification of AGN.

There are a number of unanswered questions, however, and in the
remainder of this paper I will focus on two of them: (1) Are
``molecular tori'' molecular, and (2) are they tori? As I will argue
below, in the one object currently known where the obscuration appears
to arise in a geometrically thick structure as originally envisaged,
the torus is probably atomic rather than molecular, whereas in the two
cases where we have information on the spatial distribution of
molecular gas at $r\lessapprox 1$ pc from the nucleus, the material
appears to lie in a thin warped disk. Warped accretion disks are
likely to be common in AGN, and may account for much of the
phenomenology ascribed to ``tori''.

\section{Physical Conditions in Obscuring Tori}

By definition, the inner face of the torus sees the intense radiation
field from the central source. For a given radius and source
luminosity, the gas in the torus will only be molecular if its
pressure exceeds a critical value, given by
\be
\tilde P_{\rm
cr}=P_{\rm cr}/k \simeq
1.3\times 10^{11}{L_{44}\over r_{\rm pc}^2 N_{24}^{0.9}} {\rm\; cm^{-3}\; K}
\ee 
where $10^{44}L_{44}$ erg s$^{-1}$ is the $1-100$ keV luminosity, the
distance from the luminosity source is $r_{\rm pc}$ pc, 
and $N_{\rm att}=10^{24}N_{24}\psqcm$ is the attenuating column
density between the X-ray source and the point of interest in the
torus. (The value of the column density exponent depends weakly on the
spectral shape; see Maloney, Hollenbach \& Tielens 1996 for details.)
For $\tilde P>\tilde P_{\rm cr}$, the gas is molecular, with
$T\lessapprox 10^3$ K, while for $\tilde P<\tilde P_{\rm cr}$, the gas
is atomic, warm ($T\sim 10^4$ K), and weakly ionized ($x_e\sim
0.01$). Thus, for a given X-ray luminosity and torus column density,
the torus will be atomic only if its pressure exceeds a critical
value; this pressure can be very large, even if the column density
through the torus is itself very high. (Note also that in
steady-state, the pressure in the torus must be at least equal to the
pressure of the radiation which is absorbed close to the inner face of
the torus, \ie in a region of thickness $\Delta r\ll$ the radial width
of the torus.) Equivalently, for a fixed luminosity and pressure, the
entire torus will be atomic unless the total column density exceeds a
critical value also given by equation (1). 

What diagnostics do we have for the physical conditions in the torus?
Due to the small spatial scale for $\sim$ pc-size tori, detection of
emission lines is prohibitively difficult. However, detection of
absorption lines, especially at millimeter and centimeter wavelengths
where dust extinction is not a problem, is much more promising. For
atomic tori, the best probes are (1) the 21-cm hyperfine transition of
atomic hydrogen, and (2) free-free absorption at GHz frequencies. For
molecular tori, both (1) and (2) are also potentially useful, while in
addition (3) a wide variety of molecular transitions could in
principle be seen in absorption.

Expressions for the 21-cm and free-free optical depths of tori are
given in Maloney (1996). The first detection of 21-cm absorption in a
``torus'' was toward the nucleus of the relatively nearby, luminous
radio galaxy Cygnus A (Conway \& Blanco 1995). The large linewidth
($\Delta V\approx 270\kms$) argues against this absorption arising at
large ($r\sim {\rm kpc}$) distances from the nucleus; furthermore,
such a large velocity dispersion (with random velocities comparable to
rotation velocities) is necessary to maintain a geometrically thick
torus. However, the absorption also cannot arise too close to the
nucleus: as pointed out in Maloney (1996), a torus which is too near
the luminosity source will have a large free-free optical depth at 21
cm, making the radio continuum source undetectable. This constraint
requires that the absorbing gas is at $r\gtrapprox 30-40$ pc, for
either an atomic or molecular torus. (However, this estimate ignores
the effects of stimulated emission due to the incident radio continuum
on the free-free absorption coefficient; in some cases this may be
significant.) The 21-cm absorption has now been mapped by Conway (this
volume), using the VLBA; the results indicate that the absorption
arises in a ring or thick disk with an inner radius of about 50 pc. If
the spin temperature of the hydrogen is $T\sim 10^4$ K, then the
neutral hydrogen column is comparable to that inferred from the X-ray
spectrum (Koyama 1992). In this case, although the scale is much
larger than originally envisaged, it appears that the nuclear
obscuration does arise in a geometrically thick torus.

Searches for molecular absorption from the Cygnus A torus have proven
unsuccessful (Barvainis \& Antonucci 1994, 1996; Conway \& Blanco
1995). The simplest explanation for this is that the torus is not
molecular; if the torus pressure is less that the critical pressure,
the X-ray ionization and heating rates will be high enough to maintain
the entire column in a warm atomic state. However, the observed HI
absorption could also arise in a molecular torus: the molecular gas
could be too warm or turbulent, or the covering factor of clouds at a
single velocity too small, for absorption to be detectable. An
additional, interesting possibility, first suggested by M.J. Rees (see
Barvainis \& Antonucci 1994) is that radiative excitation of molecules
by the nonthermal radio continuum is important. This possibility was
explored in detail for CO molecules in a torus in Cygnus A by Maloney,
Begelman \& Rees (1994). The basic idea is quite simple: the incident
radio continuum attempts to drive the excitation temperatures
characterizing the rotational levels to the brightness temperature
$T_b$ of the radiation field at the relevant frequency; generally
$T_b\gg T_k$, the gas kinetic temperature. This acts to reduce the
optical depth to absorption in two ways, by increasing the partition
function (spreading the population of the rotational levels over a
larger number of states) and by reducing the absorption coefficient
through the correction for stimulated emission. Including the effect
of radiative excitation, the excitation temperature of a rotational
level is given approximately by
\be
T_{\rm ex}\approx T_k\left(1+\gamma\right)
\ee
where $\gamma$ is the ratio of radiative to collisional rates. Hence,
if $\gamma$ is large, the excitation temperature can be much larger
than the gas kinetic temperature, greatly reducing the absorption
optical depth.

Maloney \etal (1994) showed that this process could plausibly be
important in the Cygnus A torus. A more detailed investigation of
radiative excitation has been carried out by Black (1998), including
non-LTE effects and examining the influence of the nonthermal
continuum on the excitation of
additional species such as OH, H$_2$CO and HCN. Radiative excitation
by the nonthermal radio continuum is quite likely to be important in
the class of radio sources known as ``Compact Symmetric Sources''
(Readhead \etal 1995), because of their large ratio of radio to X-ray
luminosities. However, the simplest explanation for the absence of any
molecular absorption is that the ``torus'' is atomic, rather than
molecular. VLBI observations of 21-cm absorption in a number of other
sources, discussed elsewhere in this volume in the contributions of
Conway and Pedlar, show the tremendous potential of this technique as
a probe of the structure and kinematics of near-nuclear gas in objects
with extended radio emission.

\section{Maser Emission and Warped Accretion Disks in AGN}

Although the obscuring material in Cygnus A appears to lie in a
geometrically thick torus as originally envisaged, albeit on
considerably larger scale, it seems likely that it is an atomic,
rather than a molecular torus. However, there are a number of AGN
where we know that there is molecular gas lying within $r\sim
1$ pc or less from the nucleus. These are the water ``megamaser''
sources, which are exclusively associated with AGN, either Seyfert 2
or LINER galaxies (see Braatz, Wilson, \& Henkel 1994, 1996; Maloney
1997). These sources have (isotropic) luminosities in the 22 GHz water
line of $L_{\rm H_2O}\sim 30-6000\lsol$; the emission is always
centered on the nucleus. Roughly $10\%$ of Seyfert 2 and LINER
galaxies show water megamaser emission (Braatz \etal 1996), and the
fact that no water megamasers are seen in Type 1 Seyferts (in which
\begin{figure}
\plotfiddle{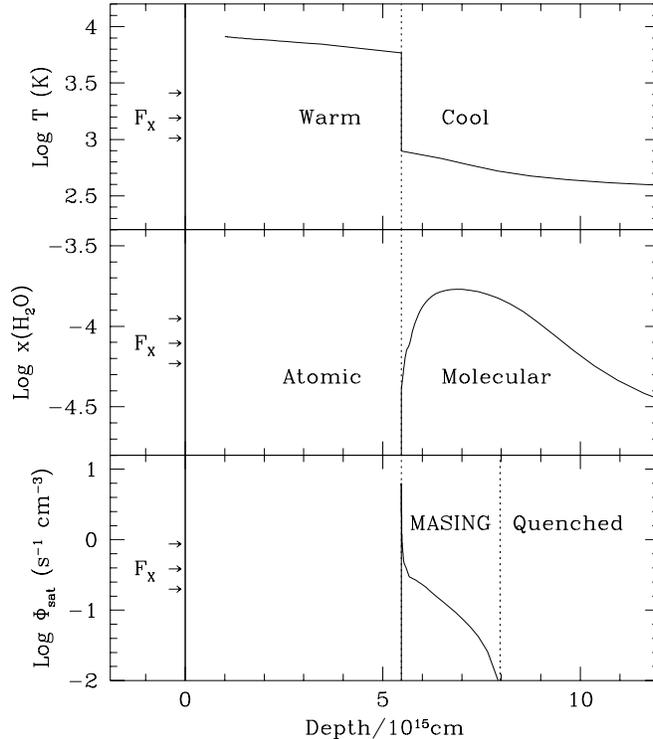}{3.6in}{0}{55}{53}{-170}{-100}
\caption{Structure of an X-ray irradiated torus; plotted as a function
of depth are the gas temperature, the water abundance, and the volume
production rate of 22 GHz maser photons when the maser emission is saturated.}
\end{figure}
our line of sight by definition lies within the opening angle of the
obscuring torus) suggests that the maser emission is directly
associated with the obscuring material. Neufeld, Maloney, \& Conger
(1994) showed that the association of water megamasers with AGN could
be understood in the context of obscuring torus models as the
consequence of irradiation of the inner face of the torus by X-rays
from the central source: as noted earlier, at some depth into the
torus (depending on the pressure) the gas will undergo a phase
transition from atomic to molecular. Across this phase transition, the
water abundance jumps from negligible values to $x_{\rm H_2O}\sim
10^{-4}$, while the temperature drops from $T\sim 10^4$ K to $T\lessapprox
10^3$ K (Figure 1). These conditions are ideal for producing maser
emission in the 22 GHz line, provided that the torus pressure $\tilde
P\lessapprox 10^{13}$ cm$^{-3}$ K, and substantial maser luminosities
($L_{\rm H_2O} \sim 10^2\lsol$ per pc$^2$ of irradiated area) are
possible. 
The maser emission is eventually quenched by photon
trapping, although photon absorption by dust may inhibit quenching, as
pointed out by Collison \& Watson (1995) (although the importance of
this process depends on the torus pressure). 

However, VLBI observations of several megamaser sources showed that
the spatial distribution of maser emission is very different than
expected from a geometrically thick torus. Mapping of the maser
emission from the LINER galaxy NGC 4258 (which provided the
first definitive evidence for the existence of a massive black hole in
a galactic nucleus [Miyoshi \etal 1995], and is discussed in detail by
Herrnstein elsewhere in this volume) revealed not the inner face of a
torus but a thin, warped disk. A similar geometry is seen in NGC 1068
(Greenhill \etal 1996; Greenhill \& Gwinn 1997), except that the disk
as traced by the maser features makes a large angle with the
radio jet axis (approximately $40^\circ$ at the outer edge of the
maser distribution); one interpretation of this is that the maser disk
in NGC 1068 is considerably more warped than that in NGC 4258. The 
maser emission is still powered by X-ray irradiation; due
to the warping of the disk, the disk is obliquely illuminated by the
central source. The maser emission can be used to derive the mass
accretion rate through the disk (Neufeld \& Maloney 1995).

There is little doubt in the case of NGC 4258 that it is the warped
accretion disk itself, as traced by the maser emission, which obscures
our line of sight to the central source. This may also be the case in
NGC 1068. Are warped accretion disks generally present in AGN, and do
they play a major role in obscuring the central source? Since the
presence of the warp is also crucial for the production of the maser
emission, as it allows the central source to illuminate the disk, the
origin of the warp is a question of considerable importance. In fact,
this problem considerably predates the discovery of the warped disk in
NGC 4258. Evidence for warped, precessing disks in X-ray binary
systems dates to the early 1970s, and the origin of these warps stood
as an unsolved theoretical problem for nearly a quarter of a century.

Recently, however, a natural mechanism for producing warps in
accretion disks has been discovered by Pringle (1996). The basis of
Pringle's instability is really quite simple, and in fact the key
feature of the instability was pointed out nearly twenty years earlier
by Petterson (1977). Consider a ring of an accretion disk that is
optically thick to both absorption and emission, orbiting a point
source of radiation. If we now warp this ring (\ie make it
non-planar), then portions of the ring will be illuminated by the
central source. The pressure exerted by the incident radiation cannot
exert any torque on the ring, since the incident flux is in the purely
radial direction. However, since the ring is optically thick to
re-emission of the absorbed radiation, the re-radiated flux will be
normal to the {\it local} plane of the disk. When integrated over the
surface of the ring, the pressure from this re-emitted radiation will
exert a non-zero torque, since the inclination (the tilt of the ring
with respect to the original rotation axis) will not be constant with
azimuth, since the ring is warped. This torque will alter the angular
momentum of the ring; in general this will lead to both precession of
the ring and a change in the ring inclination. The warp modes all have
$m=1$ symmetry, \ie they are antisymmetric.

Further work on Pringle's instability has been done by Maloney,
Begelman, \& Pringle (1996), Maloney, Begelman, \& Nowak (1997),
Maloney \& Begelman (1997), and Pringle (1997). One of the key
features of the instability is that it is an inherently {\it global}
mode: the disk twists itself up in such a way that the precession rate
is the same at every radius. The evolution of an accretion disk
subject to the instability is determined by the competition between
viscosity, which acts to flatten the disk, and radiation torque,
which acts to warp the disk. Because viscosity becomes relatively more
important compared to the radiation torque as radius decreases, there
is a minimum radius for instability: this critical radius is
approximately where the viscous and warping timescales become
equal. The actual value of the critical radius is not very sensitive
to the scaling of disk surface density $\Sigma$ with radius: for
power-law disks, with $\Sigma\propto R^{-\delta}$, the critical
radius can be written (Maloney, Begelman \& Nowak 1997) 
\be
R_{\rm cr}={1\over 2} {x_{\rm cr}^2\over \epsilon^2} R_s
\ee
where $R_s$ is the Schwarzschild radius, $\epsilon$ is the efficiency
with which rest-mass energy is converted to radiation (of order
$10\%$ for accretion onto black holes), and $x_{\rm cr}=2\pi$ 
for $\delta=3/2$ and $x_{\rm cr}\simeq 4.89\pi$ for
$\delta=-3/2$. Thus accretion disks will be unstable to
radiation-driven warping provided the disk radius exceeds a few
thousand Schwarzschild radii (assuming $\epsilon\sim 0.1$). This
condition is easily satisfied by the maser disks in NGC 4258 and NGC
1068, for which the inner edge of the maser emission occurs at
$R\simeq 3.7\times 10^4 R_s$ and $R\approx 4\times 10^5 R_s$,
respectively. 

Since, as noted above, the radiation torque becomes more
important relative to viscosity as radius increases, the warp always
grows from the outside inward. As the disk warps due to the instability,
the illumination of the disk will become non-uniform, since the finite
amplitude of the warp will cause shadowing of portions of the
disk. {\it If} the accretion disk lives long enough for the warp to
propagate to the center, then feedback between the warping and the
irradiation can occur: since the angular momentum density of the disk
increases outward, the behavior of the innermost regions of the disk
is always dominated by the advection of tilted angular momentum from
the outer portions of the disk. The change of the inner disk
orientation in turn alters the irradiation of the outer disk, as a
result of shadowing. As shown in the remarkable calculations of the
nonlinear evolution of Pringle (1997), this can lead to chaotic
behavior; this is not actually surprising, given that the governing
equation is nonlinear with a built-in time delay (the time for twist
angular momentum to propagate through the disk).

In Figure 2 I have plotted the inclination of the innermost ring of an
accretion disk under the action of Pringle's instability as a function
of time; this was calculated using the code described in Pringle
(1997). The inclination $\beta$ smoothly increases for a few hundred $t_0$,
where $t_0$ is the viscous timescale for this inner ring, until
$\beta\simeq\pi$. At this point the innermost part of the disk has
flipped completely over from its original orientation, \ie the inner
part of the disk has completely folded over. From this time
on the behavior of the inclination is chaotic. The changes in
inclination are of large amplitude, with drastic consequences for the
escape of radiation from the central source. As shown in Pringle
(1997), the obscuration of the central source by the warped
disk leads to patterns of illumination which bear a striking
resemblance to the ionization cones seen in some Seyferts, which have
usually been interpreted in the context of obscuration by a
geometrically thick torus. 

Is the source of obscuration in AGN a radiation-warped disk, rather
than a geometrically thick torus? Whether accretion disks in AGN ever
reach the chaotic state exhibited by Pringle's simulations depends on
whether they are sufficiently long-lived. The timescale for the warp
to propagate in to small radii (a few hundred times $t_0$ in the
simulation shown in Figure 2, which is typical) will be $t\sim
10^7-10^8$ years in real accretion disks. It is not clear whether
individual accretion disks actually live this long: fueling of AGN may
be dominated by episodic accretion of material (\eg molecular clouds),
in which the disk is drained by viscous accretion on a timescale
comparable to the above. However, Pringle's instability should be
generic in accretion disks around black holes, and obscuration by a
thin, warped disk alleviates many of the difficulties of maintaining a
geometrically thick torus. The simulations of Pringle (1997) are
remarkably suggestive; additional work, both theoretical and
observational, is necessary to better understand the role played by
warped disks in AGN.

\begin{figure}
\plotfiddle{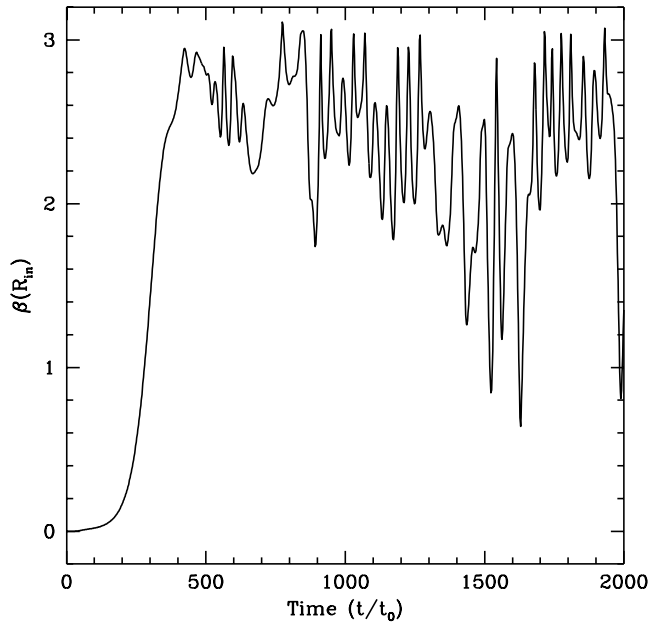}{3.2in}{0}{60}{60}{-190}{-150}
\caption{Inclination $\beta$ of the innermost ring of an accretion disk
subject to radiation torque, plotted as a function of dimensionless
time $t/t_0$.}
\end{figure}

\acknowledgements
The nonlinear warp evolution code used to produce Figure 2 was
generously provided by Jim Pringle. This research was supported by the
Astrophysical Theory Program through NASA grant NAG5-4061, and by the
NASA Long Term Space Astrophysics Program under grant NAGW-4454.

\end{document}